\documentstyle[aps,floats,epsf,psfig,twocolumn]{revtex}
\tighten
\begin{document}
\draft 
\title{Electronic Raman Scattering in Nearly Antiferromagnetic Fermi
Liquids}
\author{T.P. Devereaux$^{1}$ and A. P. Kampf$^{2}$}
\address{$^{1}$
Department of Physics, George Washington University, Washington, DC 20052}
\address{$^{2}$
Theoretische Physik III, Universit\"at Augsburg,\\
 D-86135 Augsburg, Germany}
\address{~
\parbox{14cm}{\rm 
\medskip
A theory of electronic Raman scattering in nearly antiferromagnetic
Fermi liquids is constructed using the phenomenological electron-electron
interaction introduced by Millis, Monien, and Pines. The role of "hot spots"
and their resulting signatures in the channel dependent Raman spectra is
highlighted, and different scaling regimes are addressed. 
The theory is compared to Raman spectra taken in
the normal state of overdoped Bi$_{2}$Sr$_{2}$CaCu$_{2}$O$_{8+\delta}$, 
and it is shown that many features of the
symmetry dependent spectra can be explained by the theory.}}
\maketitle

\narrowtext
It is widely believed that strong antiferromagnetic correlations are an
important ingredient needed to describe the unusual properties found in
the normal state of the cuprate superconductors. A phenomenologically
based theory proposed by Millis, Monien, and Pines\cite{MMP} 
supposes that the electronic system interacts with an independent
model spin fluctuation spectrum.
This nearly antiferromagnetic Fermi liquid (NAFL) model has been applied to 
study both spin and charge correlation properties\cite{rmp}. Of the latter,
recently Pines and Stojkovic have investigated the two-particle correlation
function in a perturbation approach
yielding the optical conductivity response\cite{optics}. In this
paper we extend their results to investigate the two-particle Raman
response.

The salient point of the calculation starts with assuming a form for
the quasiparticle-quasiparticle interaction:
\begin{equation}
V({\bf q},\Omega)=g^{2}{\alpha\xi^{2}\over{1+({\bf q-Q})^{2}\xi^{2}-i\Omega/
\omega_{sf}}},
\end{equation}
which at lowest order in the coupling constant $g^{2}$ yields for the
self energy
\begin{equation}
\Sigma({\bf k},i\omega)=-T\sum_{i\omega^{\prime}}\sum_{\bf p}
V({\bf k-p},i\omega-i\omega^{\prime})G_{0}({\bf p},i\omega^{\prime}),
\end{equation}
with $G_{0}$ the bare Green's function.
Here $\omega_{sf}$ and $\xi$ are the phenomenological,
temperature dependent spin fluctuation
energy scale and the correlation length, respectively, which can be determined
via fits to the magnetic response data\cite{rmp}. These functional 
parameters obey certain relations depending on different temperature
and doping regimes.
In the $z=1$ or pseudo-scaling regime, 
the spin correlations are strong enough to lead to changes from the
classical mean field theory $z=2$ regime. As has been pointed out
in \cite{rmp}, the $z=1$ regime is most appropriate to optimally doped
cuprates at intermediate temperatures while $z=2$ applies to optimally
doped materials at higher temperatures and/or more overdoped materials. 
For the $z=1$ scaling regime, $\omega_{sf}\xi=$constant while for
$z=2$, $\omega_{sf}\xi^{2}=$ constant. 
Once the self energy is determined, in the absence of vertex corrections
to the Raman vertex $\gamma({\bf k})$, the Raman response is simply
given by
\begin{eqnarray}
&&\chi^{\prime\prime}_{\gamma,\gamma}({\bf q}={\bf 0},\Omega)=
2\sum_{\bf k}\gamma^{2}({\bf k})
\int {dx\over{\pi}}[f(x)-f(x+\Omega)]\nonumber \\
&&\times G^{\prime\prime}({\bf k},x)
G^{\prime\prime}({\bf k},x+\Omega),
\end{eqnarray}
with $G({\bf k},x)=[x-\varepsilon({\bf k})-\Sigma({\bf k},x)]^{-1}$
and band structure $\varepsilon({\bf k})=-2t[\cos(k_{x}a)+\cos(k_{y}a)]
+4t^{\prime}\cos(k_{x}a)\cos(k_{y}a)-\mu$.
As in Ref. \cite{optics}, we use the following approximations:
(i) we neglect the real part of the self energy given in (2), (ii) we
only use the Green's function evaluated at lowest order in the coupling
constant $g^{2}$, (iii) we neglect all vertex corrections for the Raman
vertex, (iv) momentum sums are replaced by $\sum_{\bf k}=\int 
d\Omega_{\bf k}/\mid {\bf v_{\bf k}}\mid \int d\varepsilon({\bf k})$.
While (iv) does not crucially affect the results, approximations (i-iii)
while simplifying the calculations considerably miss important quasiparticle
renormalizations at larger values of the coupling. Therefore
we expect that these calculations would be most appropriate to describe the
spectra taken on overdoped cuprate superconductors. 
We believe that these approximations must be lifted in order to describe
optimally and underdoped cuprate systems. This is
analyzed in detail in a forthcoming publication\cite{forth}.

The parameters we have used for both scaling regimes are 
$g=1$eV, $\alpha=2.6$ states/eV, $t=250$meV,
$t^{\prime}/t=0.45$, filling $<n>=0.8$, $\gamma_{B_{1g}}({\bf k})=
b_{1}[\cos(k_{x}a)-\cos(k_{y}a)]$, $\gamma_{B_{2g}}({\bf k}) =
b_{2}\sin(k_{x}a)\sin(k_{y}a)$. In addition, for the $z=1$ scaling regime we
have used $\omega_{sf}\xi/a=50$meV and $a\xi^{-1}=0.1+4.64 T/2t$, while 
for $z=2$,  $\omega_{sf}\xi^{2}/a^{2}=60$meV and 
$\omega_{sf}/2t=0.0237+0.55 T/2t$. These parameters are similar to those used
in Ref. \cite{ps} to describe the Hall conductivity data in 
YBa$_{2}$Cu$_{3}$O$_{7}$. For very small incoming laser frequencies, the
Raman vertices can be written in terms of the curvature of the
electron bands. However, it has been pointed out\cite{dvz} 
that this is of questionable
use for the cuprates and therefore we have used 
the first order terms of a Brillouin zone expansion for the
vertices. Therefore the
magnitude for the scattering is arbitrary and determined by the 
dimensionless coefficients
$b_{1},b_{2}$, which are set by fitting to the data.
This has no effect on the frequency dependent lineshapes however.

Our results are summarized in Figure 1, where we have plotted the $B_{1g}$
and $B_{2g}$ Raman response for both scaling regimes. The $A_{1g}$
response is slightly more complicated due to screening effects and is
discussed in detail in Ref. \cite{forth}. The spectra for both scaling
regimes share several features. First, one immediately sees 
that the flat continuum at high frequencies
which is present in the Raman data from all
cuprate superconductors as well as several A-15 compounds is reproduced
by the theory. This is a consequence of a scattering rate 
$\Sigma^{\prime\prime}(\omega)$
which is effectively linearly dependent on $\omega$ at frequency scales
larger than $\omega_{sf}$. Moreover, the Raman response is different for
the different scattering geometries. This is due to the strong anisotropy
of the scattering rate for different directions in momentum space. In the
NAFL model, the quasiparticle scattering is strongest near ``hot spots'', i.e. 
regions of the Fermi surface which can be connected
by the antiferromagnetic wave-vector ${\bf Q}=(\pm \pi,\pm \pi)$. Since these
regions are close to the zone axes for the given band structure, this means
that the $B_{1g}$ geometry most effectively probes these hot spots while the
$B_{2g}$ geometry probes along the zone diagonals and therefore sees ``colder''
quasiparticles.  Thus the ``peak'' of the spectra is located at
larger frequencies for the $B_{1g}$ channel compared to $B_{2g}$
reflecting the larger energy scale. Lastly,
the coupling constant $g$
controls the position and character of
the peak in each scattering geometry: for smaller couplings, the peak is
more pronounced and is shifted to lower frequencies, while for larger
couplings the peak is smeared out and pushed to larger frequencies.

\begin{figure}
\hskip2.cm
\psfig{file=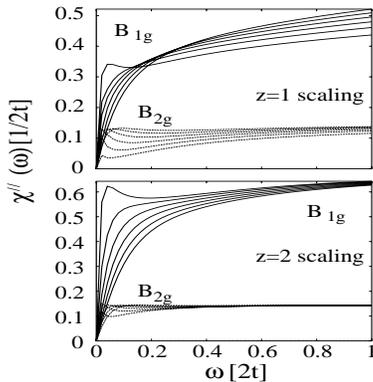,height=5.cm,width=5.cm,angle=0}
\caption[]{
Electronic Raman response for the $B_{1g}$ and $B_{2g}$ channels evaluated
at different temperatures ($T/2t=0.01,0.02,...,0.06$) for $z=1$ and $z=2$ 
scaling regimes of NAFL. Here we have set $a=b_1=b_2=1$, 
as defined in the text.}
\label{fig1}
\end{figure} 

We next discuss the results for the different scaling regimes. For these
parameter choices the differences between the two regimes are mostly
quantitative. 
In general the $z=1$ results are more strongly temperature dependent
than the results for $z=2$. This is most clearly seen in the high
frequency portion of each spectra and is due to the weaker temperature
dependence of $\xi, \omega_{sf}$ for $z=2$ scaling than for $z=1$ scaling.
In addition the $z=1$ results show stronger
differences between the results for the $B_{1g}$ and $B_{2g}$ channels
than those from $z=2$. The $B_{2g}$ spectra for both 
$z=1$ and $z=2$ as well as the $B_{1g}$ spectra for $z=2$ show a 
monotonically decreasing intensity as the temperature is increased as spectral
weight is being transfered to high frequencies. However, the $B_{1g}$ spectra
for $z=1$ show that the low frequency portion decreases with the temperature
but the high frequency portion increases with temperature. Therefore the 
spectral weight is being transfered out to only slightly higher frequencies.
This may be due to the more pronounced ``hot spots''
in the $z=1$ regime compared to the more smeared ``hot spots'' for 
$z=2$\cite{rmp}.
The $B_{1g}$ channel thus is more sensitive to the growth of the correlation
length as temperature is lowered than $B_{2g}$. 

As a consequence, the Raman spectra are quite sensitive to the details of the 
parameter
choices and therefore the shape of the Raman spectra can be a very useful
tool to understand and probe the anisotropic quasiparticle scattering
rates in much the same way as it has been used to probe the anisotropy
of the energy gap $\Delta({\bf k})$ in the superconducting state\cite{ijmpb}.

\begin{figure}
\hskip2.cm
\psfig{file=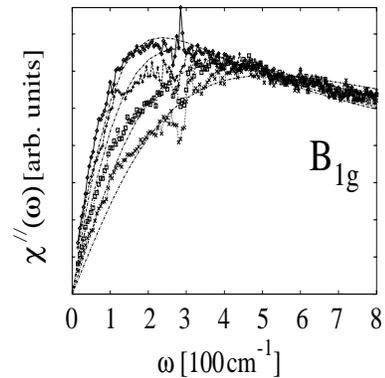,height=5.cm,width=5.cm,angle=0}
\caption[]{
Fit of the ($z=2$) NAFL theory to the $B_{1g}$ 
spectra taken in \cite{hackl} on
overdoped Bi$_{2}$Sr$_{2}$CaCu$_{2}$O$_{8+\delta}$ (T$_{c}=55 K$) at
$60K, 90K, 150K,$ and $200K$. Parameters used are given in the text.}
\label{fig2}
\end{figure} 

We now turn our attention to the comparison of the theory to the experimental
data on the cuprates. As we remarked, the approximations used in this paper
are most appropriate for systems with weak spin fluctuation scattering, and
therefore we expect that our results would best represent the data from
appreciably overdoped cuprate superconductors. Thus we consider the
results from $z=2$ scaling regime in closer detail.

As discussed in Ref. \cite{ijmpb}, it is believed that the quasiparticles in
the overdoped cuprates feel an effectively stronger impurity interaction
coming from the increased effective dimensionality of the system which
allows the quasiparticles to interact more strongly with defects residing
out of the CuO$_{2}$ plane. Therefore in addition we will consider
an isotropic impurity interaction 
$H_{imp}=\sum_{\bf k,k^{\prime}}\sum_{i,\sigma}
U e^{i({\bf k-k^{\prime}})\cdot {\bf R}_{i}}
c^{\dagger}_{{\bf k},\sigma}c_{{\bf k^{\prime}},\sigma},$
where ${\bf R}_{i}$ denotes the position of the impurity labeled by
$i$ and $U$ is the impurity potential. After averaging over the position of the
impurities, this adds 
a momentum independent term to the imaginary part of the self energy 
$\Gamma_{imp}=\pi n_{i}N_{F}\mid U\mid^{2}$, where $n_{i}$ is
the impurity concentration and $N_{F}$ is the density of states per spin
at the Fermi level.

The fit of the theory to the temperature dependent spectra for each channel 
obtained in overdoped Bi$_{2}$Sr$_{2}$CaCu$_{2}$O$_{8+\delta}$ 
(Bi 2212, T$_{c}=55 K$)
by Hackl et al. in Ref. \cite{hackl} is shown in Figs. 2 and 3
for the $B_{1g}$ and $B_{2g}$ channels, respectively. We note that
$\omega_{sf}$ and $\xi$ have not been determined via fits to magnetic
response data. These fits are thus obtained by choosing: $\omega_{sf}$
and $\xi$ at a fixed temperature, 
the magnitude of the impurity scattering rate,
a magnitude of the coupling constant $\alpha g^{2}$, and $b_{1},b_{2}$
which control the magnitude of the Raman scattering in
each channel. These parameters were chosen to fit the lowest temperature
data shown in the Figure. Then, only the temperature dependent parts of
$\omega_{sf}$ and $\xi$ were then modified to fit the data at other
temperatures as shown by the rest of the curves in Figs. 2 and 3. 
The parameters we have used are $b_{2}/b_{1}=0.417, 
\Gamma_{imp}=40$cm$^{-1}, \omega_{sf}\xi^{2}/a^{2}=13$meV, 
for all temperatures, and $\omega_{sf}=160K+0.06T[K]$. 

\begin{figure}
\hskip2.cm
\psfig{file=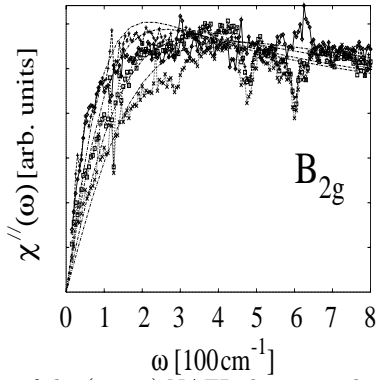,height=5.cm,width=5.cm,angle=0}
\caption[]{
Fit of the ($z=2$) NAFL theory to the $B_{2g}$ 
spectra taken in \cite{hackl} on
overdoped Bi$_{2}$Sr$_{2}$CaCu$_{2}$O$_{8+\delta}$ (T$_{c}=55 K$) at
$60K, 90K, 150K,$ and $200K$. Parameters used are given in the text.}
\label{fig3}
\end{figure} 

Further information can be obtained concerning the slope of the spectra
at vanishing frequencies. In Ref. \cite{hackl} it was shown that the
inverse of this slope has qualitatively different behaviors for different
doping regimes of various cuprate materials. Within the level of our
approximations, the low frequency Raman spectra is given by
\begin{eqnarray}
&&\chi^{\prime\prime}_{\gamma,\gamma}(\Omega \rightarrow 0)=\Omega 
\\
&&\times \int
{d\Omega_{\bf k}\over{\mid {\bf v_{k}}\mid}}\gamma^{2}({\bf k})\int
{dx\over{[2\cosh(x/2)]^{2}}}{1\over{\Sigma^{\prime\prime}({\bf k},xT)}}.
\nonumber
\end{eqnarray}

\begin{figure}
\hskip2.cm
\psfig{file=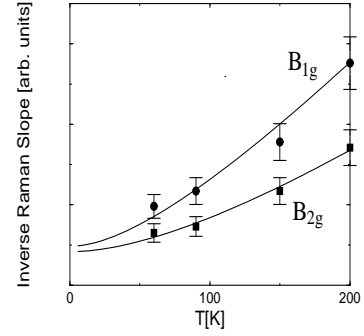,height=5.cm,width=5.cm,angle=0}
\caption[]{
Fit of the inverse slope of the Raman response for the two symmetry
channels.}
\label{fig4}
\end{figure} 

In Fig. 4 we plot 
$\Omega/\chi^{\prime\prime}_{\gamma,\gamma}(\Omega \rightarrow 0)$
obtained from the parameters used in Figs. 2 and 3 and compare the results
to the data taken in \cite{hackl}. The fit to the data is quite good
when the error bars of the data are taken into account. A 8\% uncertainty
has been included in the data\cite{opel}. We remark that similar fits
can be made with slightly different parameter choices than the ones
considered here.

In closing, we remark that while good agreement between the theory and
the data has been shown for overdoped Bi 2212, several features remain
to be explained in optimally and underdoped systems. For instance, an
almost completely temperature independent high frequency background is
observed in these systems, and the difference of the inverse slope of the 
Raman response between $B_{1g}$ and $B_{2g}$ channels grows remarkably
upon lesser dopings\cite{hackl}. Moreover, features which have been
associated with a pseudo-gap have been observed in the underdoped 
systems\cite{psuedo}. These features are discussed in greater detail in
\cite{forth}.

T.P.D. would like to acknowledge helpful conversations with D. Pines and
B. Stojkovic. Acknowledgment is made to the Donors of the Petroleum
Research Fund, administered by the American Chemical Society, for
support of this research.

\end{document}